\documentclass{article}

\usepackage{graphicx}
\usepackage{subfigure}
\usepackage{makecell}
\usepackage{xcolor}
\usepackage{float}
\usepackage{booktabs}

\usepackage{hyperref}

\usepackage[accepted]{icml2020}

\begin{document}

\twocolumn[
\icmltitle{Discover the Hidden Attack Path in Multi-domain Cyberspace \\ Based on Reinforcement Learning}
\icmlsetsymbol{equal}{*}
\begin{icmlauthorlist}
\icmlauthor{Lei Zhang}{}
\icmlauthor{Wei Bai}{}
\icmlauthor{Wei Li}{}
\icmlauthor{Shiming Xia}{}
\icmlauthor{Qibin Zheng}{}
\vspace{.1in}\\
Army Engineering University of PLA, Nanjing, China, 210007
\vspace{.1in}\\
zhangleicsphd@gmail.com, baiwei\_lgdx@126.com, 
pub\_xsm@hotmail.com, zqb1990@hotmail.com
\vspace{.1in}\\
\end{icmlauthorlist}

\vskip 0.3in
]



\printAffiliationsAndNotice{\icmlEqualContribution} 
\begin{abstract}
In this work, we present a learning-based approach to analysis cyberspace security configuration. Unlike prior methods, our approach has the ability to learn from past experience and improve over time. In particular, as we train over a greater number of agents as attackers, our method becomes better at discovering hidden attack paths for previously methods, especially in multi-domain cyberspace. To achieve these results, we pose discovering attack paths as a Reinforcement Learning (RL) problem and train an agent to discover multi-domain cyberspace attack paths. To enable our RL policy to discover more hidden attack paths and shorter attack paths, we ground representation introduction an multi-domain action select module in RL. Our objective is to discover more hidden attack paths and shorter attack paths by our proposed method, to analysis the weakness of cyberspace security configuration. At last, we designed a simulated cyberspace experimental environment to verify our proposed method, the experimental results show that our method can discover more hidden multi-domain attack paths and shorter attack paths than existing baseline methods.


\end{abstract}

\section{Introduction}

Rapid progress in AI has been enabled by remarkable advances in computer systems and hardware, but it is not widely used in cyberspace security protection. However, the intellectualization of cyberspace security protection system is an important problem facing the current cyberspace security protection \cite{DBLPSS10}. In the security management of cyberspace, actually the cyberspace should be regarded as a space composed of physical domain, digital domain, network domain and social domain, and its security protection should also be conducted as a whole \cite{DBLPnips2016}. In this process, it mainly includes the intelligence discovery of hidden attack paths, intelligent deployment of security equipment \cite{DBLPtsmc05}, intelligent monitoring of network traffic, intelligent awareness of security situation and other parts \cite{inbook009}, which comprehensively constitute the cyberspace security protection system. We believe that it is AI itself will provide the means to constitute the cyberspace security protection system, creating a symbiotic relationship between cyberspace security and AI with each fueling advances in the other.

In this work, we present a learning-based approach to discover hidden attack paths in cyberspace. Our objective is to find the shortest hidden attack path, to discover the weakest flaw in the cyberspace, thus it can effectively repair the cyberspace weaknesses and improve the security of the multi-domain cyberspace. Despite of research on this problem, it is still rely on experts to judge and evaluate the security risks existing in current cyberspace, this method is not highly intelligent and consumes a lot of human resources. Besides, the problem complexity arises from multiple domain interaction each other in cyberspace, and only digital domain or network domain can be considered in current research. In general, the cyberspace should include physical domain, cognitive domain, social domain, network domain and digital domain. Physical domain includes space and physical entity. Space includes city, campus, building, room, etc. Physical entity includes network equipment and terminal equipment, such as switch, router, computer, etc. Cognitive domain is different people have different understandings of cyberspace, resulting in different influences on cyberspace. Social domain refers to the relationship between people, for example, an attacker is a friend of the network administrator, so that it is easier to obtain higher network privileges than other attackers. Network domain is what we refer to as the traditional network space. Digital domains represent information entities used to represent digital information, such as user's names, passwords, secret keys, messages, etc. Because of multiple domain interaction each other in cyberspace, this leads to the high complexity of discovering potential attack paths. Therefore, it cannot be solved by exhaustive method to discover the shortest attack path. Even after breaking the problem into sub-problems, the state space is still larger so that the traditional methods can't to solve it.

To our knowledge, in order to realize the target, the following problems need to be solved:

\begin{itemize}
	
	\item{First, the problem of multiple domain which can interaction each other. That is the problem of multiple domain cyberspace, and how can change the existing cyberspace security risk analysis to focus on multiple domain cyberspace, enforce the pertinence and relevance of business level;}
	
	\item{Second, the problem of the alternative actions in different states is different. In traditional RL, the number of actions which agent can select is the same in any state. But in multi-domain cyberspace, the agent have different alternative actions in different states. How can solve this problem in order to make the alternative actions in different states is different.}
	
	
\end{itemize}

To address this challenge, we propose discover the multi-domain united attack paths as a Reinforcement Learning (RL) problem, where we train an agent (as an attacker) to discover the hidden shortest attack path in the multi-domain cyberspace. In each iteration of training, the attack paths are sequentially found by the RL agent. Training is guided by a fast but approximate reward signal for each of the RL agent discover the attack paths. At the same time, we proposed improved DDPG algorithm, to solve the agent can select different actions in different states.

We believe that the ability of our approach to learn from experience and improve over time unlocks new possibilities for network administrator. The experiment shows that we can achieve superior result on simulated cyberspace environment, as compared to the baselines method. Furthermore, our methods can discover the shorter attack paths comparable to human expert based method and other baseline method in same time. Although we evaluate primarily on a simulated cyberspace, our proposed method is broadly applicable to many real cyberspace.

The main contributions of this paper are as follows:
\begin{itemize}
	
	\item{We propose a unified representation method of multi-domain semantics and user's cyberspace permission, to simulate multi-domain cyberspace, which can describe the entities and entity relations in cyberspace, so that network entities in different domains can be described and expressed in a unified way, and the existing network multi-domain security state is inferred through the unified description, so as to effectively grasp the relationship between the multi-domain attacks;}
	\item{We first proposed RL on cyberspace operation and maintenance, proposed to discover the hidden attack paths by RL in multi-domain cyberspace, our experimental proved that the shorter the hidden attack path, the cyberspace security is low. It is through our proposed method that we can find the weak in cyberspace, so as to provide a reference for administrators, and improve the operation and maintenance capacity of cyberspace security;}
	\item{To address the alternative actions in different states is different, we proposed improved DDPG algorithm to solve this challenge. By introduced multi-domain action selection module, by this module, we can know an action whether or not can be executed in a state, and we can obtain a execution action ${a_t}'$ in a state. We can realize the reasonable choice of actions in different state through this method.}
	
\end{itemize}

The structure of this paper is as follows: Section 1 is introduction. In Section 2, we discuss the related works. Section 3 provides our proposed method, including the discover hidden attack paths model and improved DDPG algorithm. Section 4 presents our experiment and analysis of the experimental results. Section 5 is the conclusion.

\section{Related Work}
\subsection{Intelligent Security Protection}

Intelligent security protection mainly studied the user's behaviors and rules of user's network monitoring. In general, there is an attack in the cyberspace, when it appears, the traffic will change, we can take advantage of the attack mode type to detect cyberspace anomalies. Collect the original message of the data in the network and extract it, take the destination address and other information, establish the normal traffic model, and then use discrete wavelet transform technology analyzes and detects the data traffic to judge the cyberspace anomalies \cite{TNET902685}. 

At present, the intelligent security protection technology based on cyberspace user's action mainly relies on web data mining, user abnormal action detection and neural network based method to distinguish.

Combining traditional data mining techniques with the Internet for web mining is to extract potentially useful patterns and hidden information from web documents, web structures and service logs. Generally, according to the different objects of web mining, researcher divide web data mining into four types: web content mining, web structures mining, use record mining and web comprehensive mining \cite{Badea2015}.

In the operation process of users will retain a lot of action information, effective use of this information is the basis and key to the realization of abnormal action determination. Multi-layer log collection is implemented to support the decision of user access action, by this method, we can find the abnormal user's behavior. Using multi-level user access log, and integrate web front, user click action and URL access logic, to extract the user's access action characteristics, by a large number of calculating the average user behavior baseline characteristics, use of effective monitoring abnormal access action scoring algorithm, trace the action of the abnormal IP, the above methods can find out whether the user's behavior is abnormal or not \cite{Beutel2015}.

As an important method to deal with nonlinear systems, the neural network method has been successfully applied in the fields of pattern recognition and probability density estimation. Compared with the statistical analysis theory, the abnormal behavior analysis method based on neural network can better express the nonlinear relationship between one variable and another. The changing of abnormal network action requires the ability of behavior analysis system to analyze a large number of network packets. Moreover, many common attacks may be coordinated by multiple attackers on the cyberspace, which requires the intelligent security protection system must have the ability to deal with a large amount of nonlinear data. The method based on neural network has a fast response ability, especially for the processing of noisy data and incomplete data, so it provides a great flexibility for the analysis of intelligent security protection \cite{Kawazu2016Analytical}.

In recent years, the emergence of machine learning has made intelligent security protection become a new trend. There are many new attempts, including SVM \cite{DBLP123} \cite{Gao2017A}, K-nearest neighbors \cite{Xu2017Incremental}, Naive Bayes \cite{DBLPBK19}, random forests \cite{DBLPZH08}, neural network \cite{DBLPMK17}, deep learning and so on. The methods based on deep learning have become mainstream in the field because of their better performance. Gao proposed an model based on deep belief network, which uses a multi-layer unsupervised learning network and a supervisor-based back-propagation network \cite{DBLPQuZSQ17}. Shone used asymmetric depth self-encoders to learn network traffic characteristics in an unsupervised, not only achieved good performance on large data sets, but also reducing training time \cite{DBLPTPS18}. Yin proposed a model using RNN, compared the effectiveness of the non-depth model, and achieved good performance \cite{DBLPZFH17}. Kim proposed a model using LSTM and gradient descent strategy. The experiment result which proved the LSTM can achieve a better performance \cite{Le2017}. Sheraz conducted a comprehensive study on deep learning model, and proved that the deep learning method can not only be used in this field, but also can achieve better performance \cite{DBLPSKBHIH18}.

\subsection{Cyberspace Simulation}
Elderman et al. (2017) focus on a simulation game in cyberspace. The game is an adversarial sequential decision making problem played with two agents, the attacker and defender. The simulation cyberspace is modeled as the graph network composed of nodes where the attacker and defender move. The state of the attacker or defender is the node where they are located. Each of attacker’s actions has an attack value, and the exploit succeeds when it is larger than the defense value of the defender. The disadvantage of this method is the two agents use reinforcement learning against each other and examined their effectiveness against learning opponents. This 
research showed that we can modeled an environment to simulation the cyberspace, and the agents trying to attack the environment by reinforcement learning.

However, there are actually few environment like the previous work: in this environment, the attacker and defender will adapt and deal with each other in real-time. But according to Sharma et al. (2011), when the attackers achieved 62 of their attack goals, their attacks are detected. This means that attackers and defenders rarely compete in real time. Therefore, in this paper, we train the attack agents in the cyberspace environments that does not set the defense agents.

Both of the previous works have been validated the effects of reinforcement learning by the simulation. By this way, The second contribution of our work is to validate the effects of reinforcement learning in a simulation cyberspace environment. Our work embodies the previous works in the following points: the cyberspace configuration, the agent state, the agent actions, and the agent reward. We trained agents to learning the attack sequences steps in this simulation environment.

\subsection{Reinforcement Learning}
Reinforcement learning is commonly considered as a general machine learning model, it mainly studies how agent can learn certain strategies by interacting with the environment, to maximize long-term reward. RL is based on the Markov Decision Process(MDP) \cite{DBLPSutton98}. A MDP is a tuple $(\emph{S}, \emph{A}, \emph{T}, \emph{R}, \gamma$), which \emph{S} is the set of states and \emph{A} is the set of actions. $\emph{T}$($s_i$$\vert$$s_j$, \emph{a}): $\emph{T}\times\emph{A}$$\to$${R}$ is the reward after executing action $\emph{a}$ at stage $s_i$, and ${\gamma}$ is the discounting factor. We used $\pi$ to denote a stochastic policy, $\pi$($\emph{s},\emph{a}$): $\emph{T}\times\emph{A}$ $\to$[0,1] is the probability of executing action $\emph{a}$ at state $\emph{s}$ and $\sum_{\emph{a}\in\emph{A}}$ $\pi$($\emph{s},\emph{a}$)=1 for any $\emph{s}$. The goal of RL is to find a policy $\pi$ that maximizes the expected long-term reward. Besides, the state action value function is 

\begin{equation}\label{first_equation}
	\begin{array}{l}
		{Q}^\pi(s,a)=E\left[ \sum_{t=0}^{\infty}{\gamma}^t R(s_t,a_t)|s_0=s,a_t\sim\pi(s_t) \right] 
	\end{array}
\end{equation}

which $\gamma \in (0,1]$ measure the importance of future reward to current decisions. 

For different policies $\pi$, they represent the possibility of different actions selected in the same state, and also correspond to different rewards. A better policy can select better action in the same state, to obtain more reward.

In traditional RL, the action-value function is calculated interactively, and will eventually converge and obtain the optimal strategy, mainly including Dynamic Programming, Monte Carlo Method and Temporal-Difference Learning. After deep learning was proposed, the deep reinforcement learning method formed by combining RL is the mainstream method at present.

In the following, we introduce the mainstream RL algorithm A2C and DDPG.

\textbf{Advantage Actor Critic (A2C)}: A2C is similar to A3C (Asynchronous Advantage Actor-Critic) (Mnih et al, 2016), only there is no asynchronous part. A2C uses the advantage function instead of the raw returns in the critic network, which can be used as a measure of how good or bad the selected action values and the averages of all actions are. Both A2C and A3C are learning algorithms using the advantage. The advantage is expressed in the following equation: $A(s, a) = Q(s, a) $ - $ V(s)$. The advantage is denoted as $A(s, a)$. We denote the state of the agent as $s$ and the action as $a$. $V(s)$ represents the pure value of the state $s$ ; therefore, $A(s, a)$ represents the pure value of action $a$. According to previous studies, the advantage can stabilize the learning. A2C lacks the asynchronous part but A2C performs better than A3C. Thus, in our paper, we use A2C as a baseline, so as to stabilizes and improves the learning progress.

\textbf{Deep Deterministic Policy Gradient (DDPG)}: DDPG \cite{lillicrap2015} is a learning method that integrates deep learning neural network into Deterministic Policy Gradient(DPG) \cite{Silverarticle}. Compared with DPG, the improvement the use of neural network as policy network and \emph{Q}-network, then used deep learning to train the above neural network. DDPG has four networks: actor current network, actor target network, critic current network and critic target network. In addition to the four network, DDPG also uses experience playback, which is used to calculate the target \emph{Q}-value. In DQN, we are copying the parameters of the current \emph{Q}-network directly to the target \emph{Q}-network, that is $\theta^{Q '}=\theta^Q$, but DDPG use the following update:

\begin{equation}\label{equation5}
	\left  \{
	\begin{array}{l}
		{\theta ^{Q'}} \leftarrow \tau {\theta ^Q} + (1 - \tau ){\theta ^{Q'}}\\\\
		{\theta ^{\mu '}} \leftarrow \tau {\theta ^\mu } + (1 - \tau ){\theta ^{\mu '}}
	\end{array}
	\right.
\end{equation}

where $\tau$ is the update coefficient, which is usually set as a small value, such as 0.1 or 0.01. And this is the loss function:

\begin{equation}
	L(w) =\frac{1}{m}\sum\limits_{j=1}^m(y_j-Q(\phi(S_j),A_j,w))^2
\end{equation}

In our experiments, A2C and DDPG were compared to our proposed method so that demonstrate the superiority of our method.


\section{Methods}

In this section, we introduced our problem definition, an overview of how we formulate the problem as a reinforcement learning (RL) problem, followed by a detailed description of the reward function, action, state, reward, policy and policy updates. At last, we introduced our proposed improved DDPG algorithm to solve this problem. 
\subsection{Overview}

This section describes the proposed method, that is, the components of the training of the reinforcement learning agents. First, we define the state of the agent $s$. Second, we define the action $a$ selected by the agent in the cyberspace environment (how to select from the action lists). Third, we set rewards $r$ according to the result of the action $a$. Lastly, we introduced our method: how to improved the DDPG algorithm in this cyberspace environment. The agent accumulates the set of $s, a,$ and $ r$ observed from the learning environment as experience and proceed with learning. Figure 1 shows an overview of our method. We simulation an multi-domain cyberspace environment. Our method uses our proposed improved DDPG as the algorithm for reinforcement learning. 

\subsection{Problem Definition}
In this work, our propose is discovering hidden attack paths in multi-domain cyberspace, and the objective is to discover the shortest hidden attack paths, thus the administrators can realize the weakest in the cyberspace, so he can take measures to enhance cyberspace security. The problem is defined as given a cyberspace environment, as Figure 1 show, we stored the security information in S2. Under the current cyberspace configuration, the attacker cannot access the S2 and get the security information. Our purpose is training an attacker agent by reinforcement learning to access the S2 and get the security information. If an attacker can access S2 and to get the security information, meaning the attack is successfully. The shortest hidden attack path define: The attacker can access S2 and to get the security information and the attack sequences steps is the minimum.

We treat the problem as a MDP, that is $M = (S,A,P,R,\gamma )$, where is $s \in S$  is the current state of the cyberspace,  $a \in A$ is an attack action that is currently available, $P$ is the probability of transitions between states, $R$ is the reward value after taking an action to reach the next state. $\gamma$ is the discount factor. For the transfer probability, it can be expressed as $p(\hat s|s,a) = p({S_{t + 1}} = \hat s|{S_t} = s,{A_t} = a)$. For the reward function, it can be formally expressed as $R(s,a) = E[{R_{t + 1}}|s,a]$.

In our setting, at the initial state, $s_0$, we have a RL agent as an attacker, and an configured cyberspace. The final state $s_t$ corresponds to an attacker attack successfully or not in limited steps. At each step, the RL agent will take an action to complete an attack step. Thus, $T$ is equal to the total number of attack steps. At each step $t$, the agent begins in state $s_t$, take an action $a_t$, arrives at a new state $s_ {t+1}$, and receives a reward $r_t$ from the cyberspace environment.

We define $s_t$ to be a concatenation of features representing the state at time $t$, including where the agent stand, which computer the agent operating, which services the agent can access and so on. We will describe it in detail in Section 3.4.

$s_{t+1}$ is the next state, which includes an updated representation containing information about the newly attacker location or the attacker obtain the newly service information.

The action space is all valid actions of the state $s_t$, which is the agent will take in state $s_t$. Action $a_t$ is the agent states space of the $s_t$ that was chosen by the agent.  We will describe it in detail in Section 3.5.

Our goal in this work is to discover the shortest attack path, subject to constraints on cyberspace security configuration or security equipment. Our final reward is the a fixed value divided by number of attack steps. We will describe it in detail in Section 3.6.

\subsection{Definition of Multi-domain Cyberspace}
With the deepening of the understanding of the concept of network, especially the proposal of the concept of cyberspace, more and more scholars realize the cyberspace is affected by multi-domain behavior. Cyberspace should be defined as integrated into multiple domains such as physical domain, information domain, network domain and social domain, and takes the interconnected information technology infrastructure network as the platform, transmits signals and information through radio and cable channels, and controls the actions of entities, with special emphasis on the multi-domain attribute of the cyberspace. In this paper, because social domain mainly involves the social relationship between network administrator and attacker, the social domain attribute of multi-domain cyberspace is not discussed in this paper, we define the multi-domain cyberspace have physical domain, information domain and network domain. The physical domain mainly describes the spatial information of the equipment, the network domain mainly describes the interface, path and action related to network transmission, and the information domain mainly describes the digital information in the cyberspace.

Besides, we have some security rules in this cyberspace, including physical domain security protection rules, network domain security protection rules and information domain security protection rules. Physical domain security protection rules mainly describe the methods to prevent illegal access in the physical domain and to prevent illegal personnel from entering a certain space. The network domain security protection rules mainly describe the methods to prevent illegal access in the network domain. In general, access control lists(ACLs), static routing, VLAN partition and other methods can be used to achieve network isolation. In this paper, we focus on ACLs, which are described as allowing data to flow through a port at a source address, a port at a destination address, and a service at a destination address. The information domain security protection rules mainly describe the methods to prevent illegal access in the information domain. The main method is to encrypt the information when it is stored or transmitted. Whether encryption is through symmetric cryptography or public key cryptography, a secret key is needed to decrypt the file.

We take a deep reinforcement learning approach to the discovering the hidden attack paths problem, where an RL agent (policy network) discover the attack paths in multi-domain cyberspace; once the RL agent attack the service successfully, we will give the RL agent an positive reward, and the final reward is the reward divided by number of attack sequences steps. If the RL agent failure to attack the service or exceeds the limited number of attack steps, we will give the RL agent an negative reward. RL problems can be formulated as MDPs, consisting of three key elements: states, actions and reward. We define the following elements respectively.

\subsection{Definition of Agent State}

States: the set of possible states of the multi-domain cyberspace, and the states include the RL agent's location, the device he is operating on, and the device permissions he has. The core of cyberspace states is attacker's permissions which he can obtain by series actions. In this paper, we discuss 9 types of attacker's permissions: Space-Enter, Object-Use, Object-Dominate, Port-Use, Port-Dominate, Service-Reach, Service-Dominate, File-Dominate, and Information-Know, respectively. Space-Enter, Object-Use and Object-Dominate is a physical domain's permission, the Space-Enter is meaning attacker enters a physical space, Object-Use is meaning attackers have permission to use a device, terminal or equipment, Object-Dominate meaning attackers can dominate an equipment, Object-Use and Object-Dominate are the differences between the former only can use the equipment with the current state of the equipment, while the latter can change the configuration of equipment. Port-Use, Port-Dominate, Service-Reach are network domain's permission, Port-Use is meaning the attacker can use the port to access network, Port-Dominate is meaning the attackers can change the state or configuration of a port, Service-Reach is meaning the service request information flow to achieve the service, but cannot use the service. File-Dominate is meaning the attacker can through the secure authentication service, to use the service. File-Dominate, and Information-Know are digital domain's permission. File-Dominate is meaning the attacker can read, delete and modify the file or even more. Information-Know is meaning the attacker obtain the security information, for example, administrator user name, administrator password, administrator key or so on.

In this paper, we set the attacker's cyberspace permission to the RL states, We set a state list, including the attacker's devices permissions in cyberspace. For example, if an attackers in the outer space, we set the attacker's  state as follows: Room A Space-Enter is 0, Room B Space-Enter is 0, Terminal A Object-Use is 0, Terminal A Object-Dominate is 0, Port B Port-Use is 0, Port B Port-Dominat is 0, et al. And then, the attacker have an action: enter room A, the the state set as Room A Space-Enter is 1, and other state value is also 0. When the attacker have an action, the states is also changing because the attacker's action.

\subsection{Definition of Agent Action}
Actions: the set of actions that can be taken by the agent and can change the cyberspace environment's states (e.g., enter a room, operate or control a computer, access a service by a port).

But in this cyberspace environment, action space is very large, and the attacker's actions is limited in a state, for example, if the attacker's state is in the outer space, he only have the enter-room or stay still two actions to choose, he don't have the operate a computer, control a computer or other actions, this is different from standard RL architecture, therefore, we will give some constraints in states when the attacker select actions, and we'll cover this in more detail in Section 3.7.

\subsection{Definition of Agent Reward}
Reward: the reward for agent taking an action in a state. 

This section describes the reward setting. In our experiment of Section 4, the goal of the agent is to obtain the security information in S2 and obtain the key to decrypt the information. In that case, the first compromised is that the attacker how can obtain the Object-Use and Object-Dominate of S2, so as to get the password to decrypt the security information, that is the attacker obtain the Information-Know permission of the security information, it is means the attack is successfully. Because the multi-domain security rules, the attacker cannot access the S2 directily, in order to access the S2, the attacker will access the FW2 to modify the ACLs, to allow the attacker can access the server S1 and S2, then the attacker can access S2, if he obtain the security information successfully, he also back to modify the ACLs to restore the original state. Therefore, reward $r$ setting is as follows:
\begin{itemize}
	\item $r$=500000/$t$ if the attacker obtain the Information-Know of security information in S2, the $t$ is the attacker's successfully sequences steps;
	\item $r$=10 if the attacker obtain the File-Dominate of the security information in S2;
	\item $r$=5 if the attacker obtain the Object-Dominate and Service-Dominate of S2;
	\item $r$=5 if the attacker obtain the Object-Dominate and Service-Dominate of S1;
	\item $r$=1 if the attacker obtain the Port-Use of S1;
	\item $r$=1 if the attacker obtain the Port-Use of S2;
	\item $r$=-0.1 if the attacker have an action but obtain no permissions.
\end{itemize}
Reward $r$=-0.1 means punishment for the agents. The agents try to maximize the reward. Therefore, if $r$=-0.1 is set for each action, the agents try to reach the goal as soon as possible. This corresponds to a situation where attacker achieve the goal as soon as possible to minimize attack sequences steps. Besides, the episode ends when the agent attack successfully and the attack sequences exceeds 10000 steps.

\subsection{Improved DDPG Algorithm}

The main function of our model is to discover the hidden attack path by a RL agent under a certain cyberspace configuration.

The model takes DDPG algorithm, we are using an agent to represent the attacker. In the process of discovering the optimal hidden attack path, the agent first selects the action in the current state, which can change the environment and the agent's state. At the same time the agent will obtain certain reward or negative reward. Besides, the change in the agent's state enables it to perform other actions to obtain more rewards. As a result, the agent discover the hidden attack path in this cyberspace configuration by trial and error. The model is shown in Figure \ref{figure2222}.

The process of to find the shortest attack sequences steps to select the corresponding action \textbf{\emph{a}} according to the current state \textbf{\emph{s}} of the cyberspace, it also means find the corresponding policy mapping function $R(s) \to A$, to make the long-term reward of agent maximum. In this process, the policy can be divided into two categories, namely the deterministic policy and the stochastic policy, deterministic policy is for the state, the conviction of corresponding output(action). In general, the deterministic policy algorithm efficiency is high, but the lack of ability to explore and improve. Rather than the deterministic policy, the stochastic policy join corresponding random value, enables the stochastic policy to have certain ability of exploration. For our model, since the action value range is generally not large in practical problems, a deterministic policy is adopted to ensure better performance of the model.

The model is a standard RL model, through the study of the awareness of environment, the agent will take the action and get a reward, the goal of the agent is to maximize rewards, and then to further training of the agent. In this paper, we will to take DDPG algorithm and improved it, its main architecture as shown in Figure \ref{figure11}.

With the standard DDPG algorithm, it also consists of four networks and one experience replay memory. Among them, the experience replay memory is mainly responsible for storing the state transfer process of $ <s,a,r,s'> $, then, by means of small batch sampling, the corresponding transferred samples are extracted to train the corresponding neural network so as to avoid the strong correlation between samples. Among the four networks, there are two policy networks(Actor) and two \emph{Q}-networks(Critic), namely the online policy network, the target policy network, the online \emph{Q}-network and the target \emph{Q}-network. The policy network mainly simulates the attacker's policy through the deep neural network, which takes the current state as the input and the output as the corresponding action. The \emph{Q}-network is mainly used to estimate the expectation of the final reward value obtained if the policy is continuously executed after the current action is executed in a certain state. The input is the current state, the current action and the output is the \emph{Q}-value. If only a single neural network is used to simulate policy or \emph{Q}-value, the learning process is unstable. So in DDPG algorithm respectively, policy network and \emph{Q}-value network create copies of two networks, two networks are known as the online network, two networks are known as the target network, online network is the current training of network, the target network is used to calculate the training goal, and after a short period of time, the model of online networks parameter updates to the target networks, so as to make the training process is stable, easy to convergence.

We have improved the standard DDPG algorithm, which is different from the standard DDPG algorithm in two aspects:

\begin{itemize}
	\item  In improved DDPG algorithm, we introduced the multi-domain action selection module.
	
	Different from standard DDPG algorithm, the biggest change is that the introduction of multi-domain action selection module. In the standard DDPG, the actions which agent can choose in each state is the same. But in this environment, when the attacker select the attack paths in cyberspace, he have different alternative actions in each state. For example, if the attacker's state is in the outer space, he only have the action 'enter room' or 'keep still', but in standard DDPG algorithm, he have all the actions to select, obviously, when attacker in outer space, he cannot have the action 'control terminal'. In order to make DDPG algorithm can choose different actions in different states, joined the multi-domain action selection module. This module's input is online policy network's output, we called this is theory action $a_t$, then a linear change under this current state, and perform the actual action $a_t'$, the actual execution action $a_t'$ into multi-domain action execution module, get the corresponding reward $r_t'$. In the end, the corresponding actual execution of action $a_t'$ and the corresponding reward $r_t'$ return online policy network. Through this method, the reasonable choice of actions in different states can be realized.
	
	\item Second, the input of experience playback memory is different.
	
	In order to ensure that the multi-domain action selection conforms to the constraints of the actions on the state, the input of the experience playback memory is increased, not only by the online policy network to store the sequence $({s_t},{a_t},{r_t},{s_{t + 1}})$, that is, execute the action $a_t$ in the state $s_t$, get the reward value $r_t$, and convert to the next state to $s_{t+1}$. Moreover, since the corresponding relationship between the state and the action needs to be considered when selecting the action, it is avoided that the policy network chooses the action that is not feasible in the state. Therefore, when the policy network selects an inoperable action $a_t$ in the state $s_t$, it is not only necessary for the multi-domain action selection module to use a linear transformation to map it to a feasible action $a_t'$, in addition, relevant action sequences $({s_t},{a_t},{- \infty},{s_{t + 1}})$ need to be taken to indicate that actions $a_t$ are executed in the state $s_t$, and the subsequent state obtained is still $s_t$, and the reward at this time is a huge negative value, so as to ensure that relevant actions are not selected in the process of training the policy network.
\end{itemize}

In terms of network architecture, the two policy networks have the same architecture, whose input is the state of network and output is the action to be selected. Structurally, a RNN hidden node is added between the original DDPG input layer and the hidden layer. The transformed policy network is divided into 5 layers. The first layer is the input layer; The second layer is the RNN hidden layer, which contains 32 GRU nodes. The third layer and the fourth layer are the full connected layer, including 48 full connected nodes. The activation function uses the ReLu function. The fifth layer is the output layer, use the sigmoid function as the activation function, and finally output a multi-dimensional vector representing the multi-domain action that needs to be performed.

In addition, the two \emph{Q}-networks have another architecture, whose input is not only the state of the network, but also includes a multidimensional vector, representing the corresponding multi-domain actions, and the output is a scalar. The network is divided into four layers. The first layer is the input layer; The second layer and the third layer respectively contain 48 fully connected nodes. The activation function uses the ReLu function. The fourth layer is the output layer, which outputs a scalar and uses the linear function as the activation function, representing the \emph{Q}-value of the corresponding state and action.

\section{Experiments}

\subsection{Experiment Environment}

\begin{figure}[h]
	
	\centering
	\includegraphics[height=4.2cm, width=6.8cm]{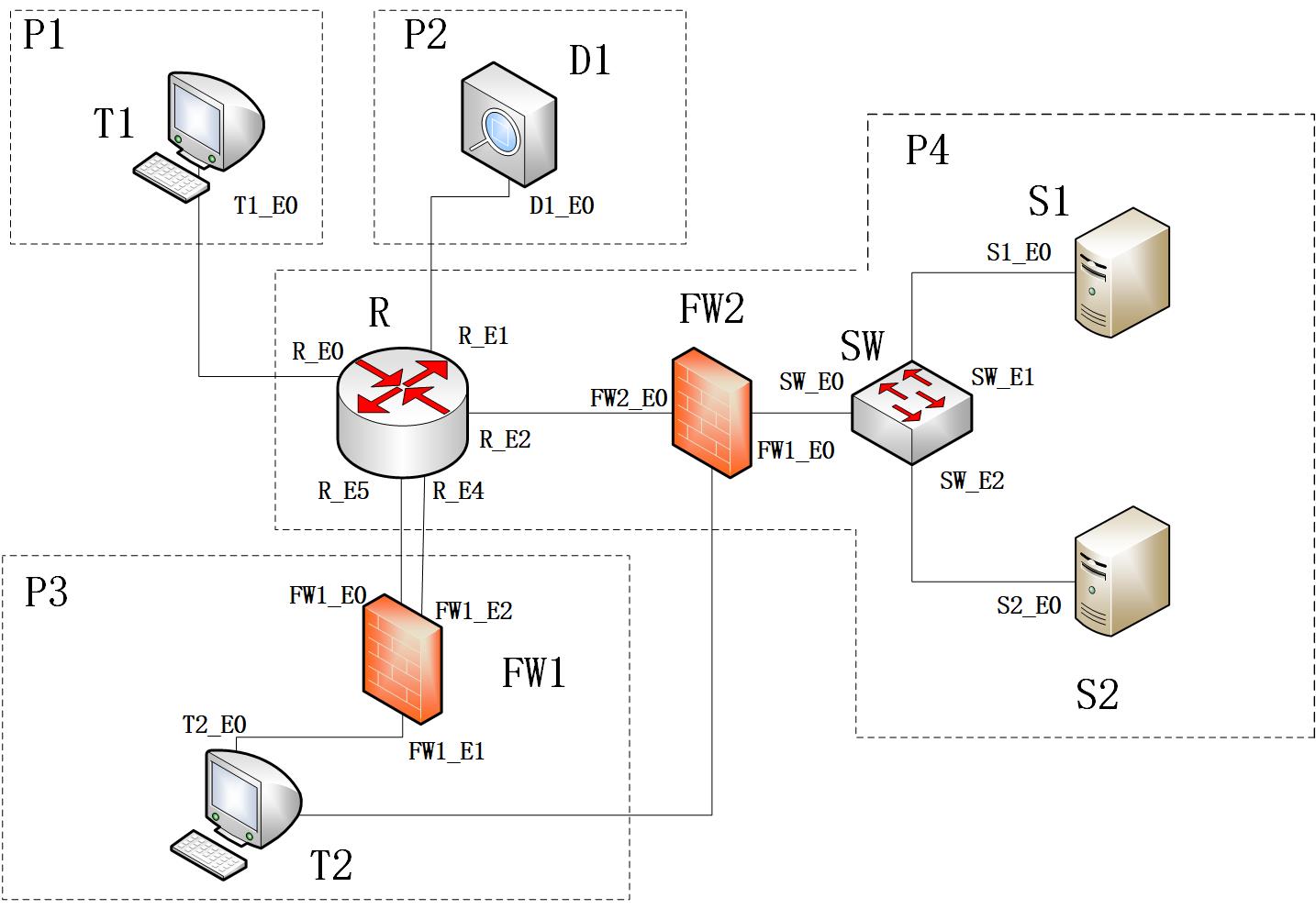}
	
	\caption{Experiment Environment Topology}
	
	\label{figure336}
	
\end{figure}

In this experiment, we modeled an cyberspace environment as our experiment data, to verify the effectiveness of our method. In this experiment environment, there are five spaces in total. The outermost space is the whole physical space, representing a region. P1 is the region where terminal located, P2 is the region where VPN equipment located, P3 is the location of the communication team, and P4 is the communication hub. There are 5 kinds of equipment, including computer(T1 and T2, respectively stored in P1 and P3), firewall(FW1, FW2, respectively stored in P3 and P4), sensor (D1, stored in P2), router (R, stored in P4) and switch (SW, stored in P4), server (S1, S2, stored in P4) and its equipment connection relationship as shown in Fig. \ref{figure336}. The security information is stored in S2. If an attacker can access S2 to get the security file and obtain its 'Information-Known' permission, meaning the attack is successfully. Also, our purpose is not only attack successfully, but also the attack sequences is shortest.


In this network, there are 15 network services, as shown in Table \ref{tab111}

\begin{table*}
	\caption{Network Services in the Experiment Environment}
	\label{tab111}
	\begin{tabular}{ccccc}
		\toprule
		Web Service & Web Services's Role & Service Support Equipment & Service 
		Dependent Port & Service Password\\
		\midrule
		T2\_manager & Remote Management equipment & T2 & T2\_E0 & None\\
		FW1\_manager & Remote management equipment & FW1 & FW1\_E2 & 
		FW1\_password\\
		FW2\_manager & Remote management equipment & FW2 & FW2\_E2 & 
		FW2\_password\\
		S1\_web & Web services in server S1 & S1 & S1\_E0 & S1\_web\_password\\
		S2\_web & Web services in server S2 & S2 & S2\_E0 & S2\_web\_password\\
		\bottomrule
	\end{tabular}
\end{table*}

In this environment, because of the firewall FW1 equipment are in need of 
remote management, FW1-password remains in FW1, at the same time, due to the T2 
maintains FW2 and S1, so T2 store password FW2\_password and S1\_web\_password, 
in this environment, to ban other flow of information. But we can know, an 
attacker can through the multiple domain joint attack, which can obtain the 
security information stored on the server S2, a possible attack path is as 
follows:

First, attacker enter space P2 and obtain the management service password 
FW1\_password of firewall FW1;

Second, use device T1 or D1 to access the management service of FW1, add access 
control list: allow T1 or D1 to access the management service of T2, that is 
T2\_manager;

Third, get the password FW2\_password of firewall FW2 stored on T2 and the 
password S1\_web\_password of service S1\_web through T2\_manager;

Fourth, use T2\_S1 port, access firewall FW2\_manager, add access control list: 
allow T1 or D1 access service S1\_web and S2\_web;

Fifth, use T1 or D1 to access the service S1\_web and get the password 
S2\_web\_password of S2\_web, at this point, the attacker's higher permissions 
have been obtained.

At last, the attacker can use T1 or D1 to access the service S2\_web to get the 
security information by the S2\_web\_password.

In this process, three key firewall security policy changes are involved: on 
firewall FW1, T1 or D1 are allowed to access T2's management service 
T2\_manager; On firewall FW2, allow T1 or D1 access to service S1\_web; On 
firewall FW2, allow T1 or D1 access to the service S2\_web.

\subsection{Experiment Process}

During the experiment, an agent (attacker) is introduced, located in the outermost space, and then, in the cyberspace environment shown in Figure \ref{figure336}.

Each episode's termination condition is the attack sequence steps exceeds 10000. And we train 500 episode. If an attacker attack successfully, the attack sequence steps $n$ will be recorded, and he will go back to the outermost space, start looking for the attack sequences again, until the number of  attack sequences steps exceeds 10000 in an episode.

\begin{figure*}[h]
	\centering
	\subfigure[IDDPG Method]{\includegraphics[width=8cm]{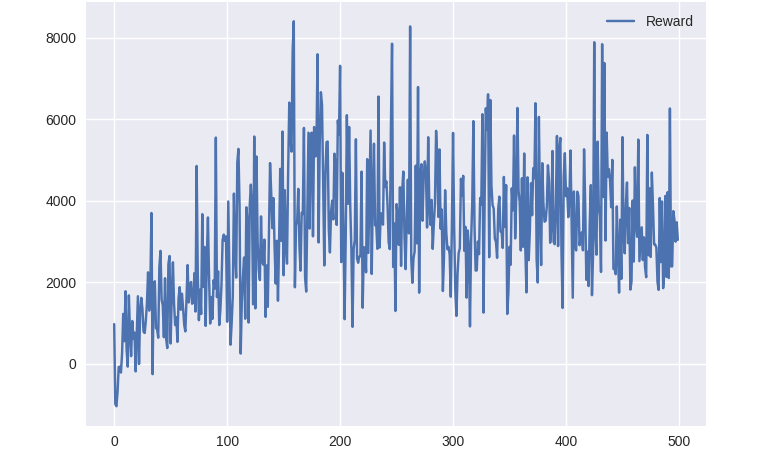}}
	\subfigure[DDPG  Method]{\includegraphics[width=8cm]{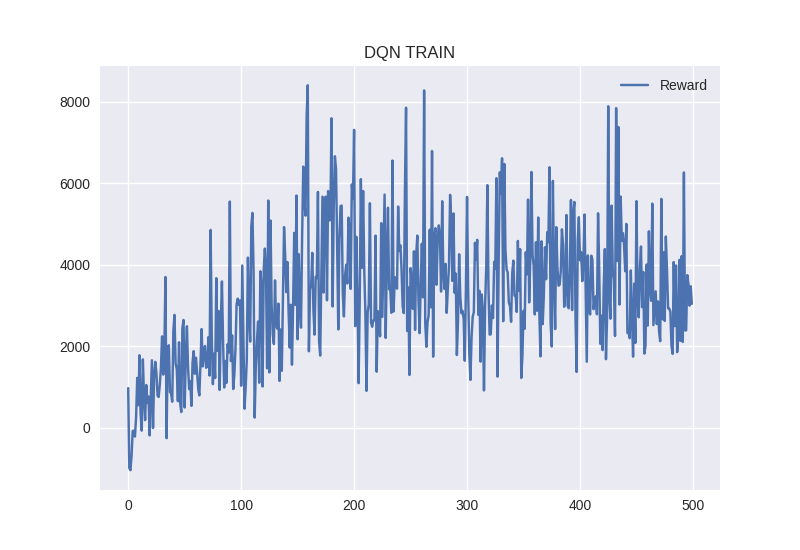}}
	\subfigure[A2C Method]{\includegraphics[width=8cm]{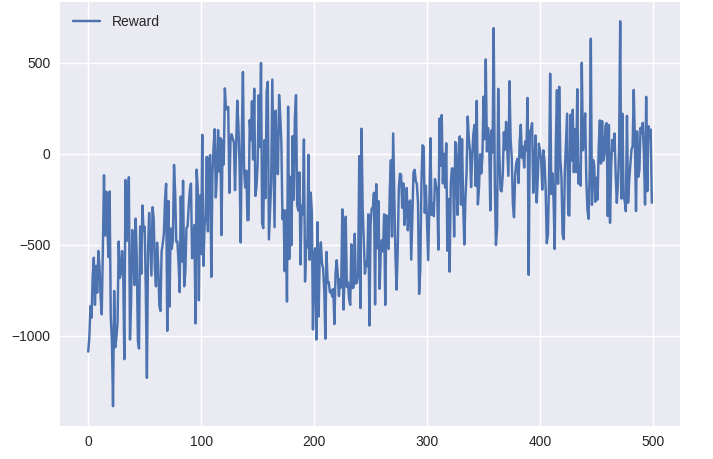}}
	\subfigure[DQN Method]{\includegraphics[width=8cm]{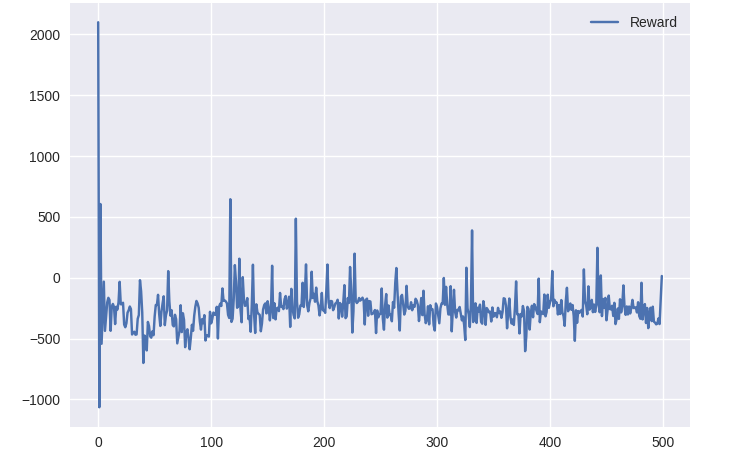}}
	\caption{ Different Methods Reward Results }
	\label{fig4}
\end{figure*}
According to the RL model and improved DDPG algorithm, define the corresponding state, action, reward, etc. The relevant settings are as follows:

On the set of state, we set a length of 106 vector to represent a state of different position on the value of the attacker, respectively from the spaces, the ports, services or information, in setting a state vector, for example, if the attacker is in a certain space, the value representing that space is set to 1, otherwise it is 0. If the attacker can use a port, set the value representing the port to 1, otherwise it is 0; If the attacker is connected to a service, set the value representing the service to 1, otherwise 0; If the attacker obtain security information, set the value representing that information to 1, otherwise it is 0.

Attackers have different action in different states. We have introduced in Section 3.5. For example, when the attacker can dominate the management service FW1\_manage of FW1 in the current state, he can add the corresponding access control list for firewall FW1 in the current state. Otherwise, he cannot add the access control list for FW1.

In terms of the reward setting, different reward are set for the attacker according to the degree of the completion of the attack path. Among them, in an episode, the reward is set in Section 3.6.

\subsection{Baseline}
In order to verify the effectiveness of our proposed method, we propose three comparative RL methods to verify the effectiveness of our proposed method:

DQN method: DQN is a RL standard algorithm, DQN uses a neural
network to predict Q-value and constantly updates the neural
network to learn the shortest attack path. There are two neural networks in DQN. One is a relatively fixed network, called target network, is used to obtain the value of Q-target, the other is evaluate-network, used to get the value of Q-evaluate.

A2C method: A2C is a RL standard algorithm, we have introduce this method in section 2.3. This method similar to A3C, but there is no asynchronous part. 

DDPG method: DDPG is also a RL standard algorithm, we have introduce this method in section 2.3.

Improved DDPG method(IDDPG): this method is this paper proposed.

\subsection{Results and Discussions}

\begin{table}[!htbp]
	\caption{Attack Sequences Steps in Different Security Rules}
	\label{extremos456}
	\centering
	\begin{tabular*}{\hsize}{ p{2cm} p{1.2cm} p{1.2cm} p{1.2cm} p{1.2cm}}
		\hline
		security rules  & 3 & 4 & 5 & 6 \\
		\hline
		DQN & 3533  & 4323 & 7231 & 8313 \\
		A2C & 3222 & 4213 & 7332 & 8812 \\
		DDPG & 2434 & 4765 & 7240 & 8551 \\
		IDDPG & 995 & 3322 & 6321 & 7233 \\
		
		\hline
	\end{tabular*}
\end{table}
Under the above conditions, learn 500 episodes in environments, and we recorded these results: 

Result A: We added some security rules, to compare the average successfully attack sequences steps, that is, we added up all the successfully attack sequences steps and divided by the number of successfully attack.

From Tabla \ref{extremos456}, we added security rules, respectively the number of security rules is 3, 4, 5, 6. In general, when the more security rules there are, the attack sequences steps is longer, even so the attacker cannot attack successfully. From the Result A, Tabla \ref{extremos456}, it also confirm the opinion. As we can see, as the number of security rules increases, the attack sequences steps get longer and longer in all methods. Therefore we can confirm, if the cyberspace have no security rules, the attackers more easily to the obtain their target, this also proved that the important of the cyberspace configure. Result A shows the we can use the attack sequence steps to assess the level of cyberspace security. That is, if the attack successfully sequences is shorter, the cyberspace security is low, that is the cyberspace configuration is bad. On the contrary, if the attack sequences steps is longer, that is the cyberspace configuration is better. Therefore, in our experiment, We can use the attack successfully sequences steps to represent the cyberspace security.

Following the experiment, we use the number of security rules is 3, to verify our proposed method is useful and efficient. 

Result B: The reward the attacker obtain.

We can see the Result B form Figure 5, first of all, with the increasing the number of training, \emph{R} present a slow upward trend, until finally tend to be convergent, it is in a learning process of RL, shows that the proposed model conformed to the characteristics of reinforcement learning, can discover the hidden attack paths in the cyberspace environment, and discover the shorter hidden attack paths constantly, so as to verify the effectiveness of the proposed model is presented in this paper.

Secondly, Figure 5 represent four methods obtain the reward, we can see our proposed method IDDPG have the highest reward than other methods; This may account for the following reasons: our proposed method has more successfully attacks, and it has shorter attack sequences steps.

We define the Attack Successfully Number is the attacker attack successfully in an experiment, the Minimum Steps is the in an episode the shortest attack successfully sequences steps.

Result C: The Attack Successfully Number; 

Result D: The Minimum Steps;

Result C and D is recorded in Tabla 3.

\begin{table}[htbp]
	\caption{Experiment Results}
	\label{extremos45}
	\begin{tabular*}{\hsize}{lcc}
		\hline
		& Attack Successfully Number & Minimum Steps  \\
		\hline
		DQN &  1203 & 1800   \\
		A2C & 2533 & 2400   \\
		DDPG & 3711 & 2100   \\
		IDDPG & 6705 & 750   \\
		\hline
	\end{tabular*}
\end{table}

Thirdly, we can see the Table 3, in Result C and Result D, our proposed method have the minimum attack sequences, it is shown that our proposed method can discover the shortest attack sequences steps under the same conditions, also, the attack successfully number is also maximum. From the experiment results, our proposed method is superiority than existing methods. 

From the experiment result, first we confirm the cyberspace configuration can affect the cyberspace's security, then we can use the attack sequences steps as the measure of the cyberspace security; Following, we use the different method in the same cyberspace environment, we compared the reward, the Attack Successfully Number, the Minimum Steps, our proposed method have the higher reward in the experiment, and the most Attack Successfully Number, the minimum Minimum Steps, from this experiment results, we can conclude our proposed method can attack more successfully, to get more reward in an episode. As a result, the attacker will have shortest steps, will discover the shortest attack sequences steps in cyberspace, then will provide reference to the administrator where the weakest in the cyberspace. In a word, our proposed method is superiority than existing methods.

\section{Conclusion}
In order to discover the vulnerability of the cyberspace, we proposed a learning-based approach to discover the hidden attack path. Meanwhile, we has been learn about the cyberspace weakness metrics, that is the attacker attack successfully sequences steps, and finally has carried on the experimental verification. This method can comprehensively consider the mutual influence of the multiple domain configuration in the cyberspace, and can take an intelligent method to analysis the weakness of the cyberspace, which has a strong practical value.

This paper analysis a typical cyberspace environment and applies the reinforcement learning method to discover the hidden attack paths in configured cyberspace, which has achieved better results. However, the cyberspace environment in this paper is limited. In the next step, we hope to apply the reinforcement learning to more cyberspace operation and maintenance management, and achieve better results.


\bibliography{arxivpaper}


\begin{thebibliography}{22}


\ifx \showCODEN    \undefined \def \showCODEN     #1{\unskip}     \fi
\ifx \showDOI      \undefined \def \showDOI       #1{#1}\fi
\ifx \showISBNx    \undefined \def \showISBNx     #1{\unskip}     \fi
\ifx \showISBNxiii \undefined \def \showISBNxiii  #1{\unskip}     \fi
\ifx \showISSN     \undefined \def \showISSN      #1{\unskip}     \fi
\ifx \showLCCN     \undefined \def \showLCCN      #1{\unskip}     \fi
\ifx \shownote     \undefined \def \shownote      #1{#1}          \fi
\ifx \showarticletitle \undefined \def \showarticletitle #1{#1}   \fi
\ifx \showURL      \undefined \def \showURL       {\relax}        \fi
\providecommand\bibfield[2]{#2}
\providecommand\bibinfo[2]{#2}
\providecommand\natexlab[1]{#1}
\providecommand\showeprint[2][]{arXiv:#2}

\bibitem[\protect\citeauthoryear{Akashdeep, Manzoor, and Kumar}{Akashdeep
  et~al\mbox{.}}{2017}]%
        {DBLPMK17}
\bibfield{author}{\bibinfo{person}{Akashdeep}, \bibinfo{person}{Ishfaq
  Manzoor}, {and} \bibinfo{person}{Neeraj Kumar}.}
  \bibinfo{year}{2017}\natexlab{}.
\newblock \showarticletitle{A feature reduced intrusion detection system using
  {ANN} classifier}.
\newblock \bibinfo{journal}{\emph{Expert Syst. Appl.}}  \bibinfo{volume}{88}
  (\bibinfo{year}{2017}), \bibinfo{pages}{249--257}.
\newblock
\urldef\tempurl%
\url{https://doi.org/10.1016/j.eswa.2017.07.005}
\showDOI{\tempurl}


\bibitem[\protect\citeauthoryear{B and Muneeswaran}{B and Muneeswaran}{2019}]%
        {DBLPBK19}
\bibfield{author}{\bibinfo{person}{Selvakumar B} {and} \bibinfo{person}{K.
  Muneeswaran}.} \bibinfo{year}{2019}\natexlab{}.
\newblock \showarticletitle{Firefly algorithm based feature selection for
  network intrusion detection}.
\newblock \bibinfo{journal}{\emph{Comput. Secur.}}  \bibinfo{volume}{81}
  (\bibinfo{year}{2019}), \bibinfo{pages}{148--155}.
\newblock
\urldef\tempurl%
\url{https://doi.org/10.1016/j.cose.2018.11.005}
\showDOI{\tempurl}


\bibitem[\protect\citeauthoryear{Badea, Croitoru, and Gheorghică}{Badea
  et~al\mbox{.}}{2015}]%
        {Badea2015}
\bibfield{author}{\bibinfo{person}{Adrian Badea}, \bibinfo{person}{Victor
  Croitoru}, {and} \bibinfo{person}{Daniel Gheorghică}.}
  \bibinfo{year}{2015}\natexlab{}.
\newblock \showarticletitle{Computer networks security based on the detection
  of user's behavior}.
\newblock


\bibitem[\protect\citeauthoryear{Beutel, Akoglu, and Faloutsos}{Beutel
  et~al\mbox{.}}{2015}]%
        {Beutel2015}
\bibfield{author}{\bibinfo{person}{Alex Beutel}, \bibinfo{person}{Leman
  Akoglu}, {and} \bibinfo{person}{Christos Faloutsos}.}
  \bibinfo{year}{2015}\natexlab{}.
\newblock \showarticletitle{Graph-Based User Behavior Modeling: From Prediction
  to Fraud Detection.}
\newblock


\bibitem[\protect\citeauthoryear{Gao, Gao, Yi-Yue, and Wang}{Gao
  et~al\mbox{.}}{2017}]%
        {Gao2017A}
\bibfield{author}{\bibinfo{person}{Ni Gao}, \bibinfo{person}{Ling Gao},
  \bibinfo{person}{H.~E. Yi-Yue}, {and} \bibinfo{person}{Hai Wang}.}
  \bibinfo{year}{2017}\natexlab{}.
\newblock \showarticletitle{A Lightweight Intrusion Detection Model Based on
  Autoencoder Network with Feature Reduction}.
\newblock \bibinfo{journal}{\emph{Acta Electronica Sinica}}
  (\bibinfo{year}{2017}).
\newblock


\bibitem[\protect\citeauthoryear{Heo and Varshney}{Heo and Varshney}{2005}]%
        {DBLPtsmc05}
\bibfield{author}{\bibinfo{person}{Nojeong Heo} {and}
  \bibinfo{person}{Pramod~K. Varshney}.} \bibinfo{year}{2005}\natexlab{}.
\newblock \showarticletitle{Energy-efficient deployment of Intelligent Mobile
  sensor networks}.
\newblock \bibinfo{journal}{\emph{{IEEE} Trans. Systems, Man, and Cybernetics,
  Part {A}}} \bibinfo{volume}{35}, \bibinfo{number}{1} (\bibinfo{year}{2005}),
  \bibinfo{pages}{78--92}.
\newblock
\urldef\tempurl%
\url{https://doi.org/10.1109/TSMCA.2004.838486}
\showDOI{\tempurl}


\bibitem[\protect\citeauthoryear{Kawazu, Toriumi, Takano, Wada, and
  Fukuda}{Kawazu et~al\mbox{.}}{2016}]%
        {Kawazu2016Analytical}
\bibfield{author}{\bibinfo{person}{Hirotaka Kawazu}, \bibinfo{person}{Fujio
  Toriumi}, \bibinfo{person}{Masanori Takano}, \bibinfo{person}{Kazuya Wada},
  {and} \bibinfo{person}{Ichiro Fukuda}.} \bibinfo{year}{2016}\natexlab{}.
\newblock \showarticletitle{Analytical method of web user behavior using Hidden
  Markov Model}. In \bibinfo{booktitle}{\emph{2016 IEEE International
  Conference on Big Data (Big Data)}}.
\newblock


\bibitem[\protect\citeauthoryear{Kim and Reddy}{Kim and Reddy}{2008}]%
        {TNET902685}
\bibfield{author}{\bibinfo{person}{Seong~Soo Kim} {and}
  \bibinfo{person}{A.~L.~Narasimha Reddy}.} \bibinfo{year}{2008}\natexlab{}.
\newblock \showarticletitle{Statistical Techniques for Detecting Traffic
  Anomalies through Packet Header Data}.
\newblock \bibinfo{journal}{\emph{IEEE/ACM Trans. Netw.}} \bibinfo{volume}{16},
  \bibinfo{number}{3} (\bibinfo{date}{June} \bibinfo{year}{2008}),
  \bibinfo{pages}{562–575}.
\newblock
\showISSN{1063-6692}
\urldef\tempurl%
\url{https://doi.org/10.1109/TNET.2007.902685}
\showDOI{\tempurl}


\bibitem[\protect\citeauthoryear{Le, Kim, and Kim}{Le et~al\mbox{.}}{2017}]%
        {Le2017}
\bibfield{author}{\bibinfo{person}{Thi Thu~Huong Le}, \bibinfo{person}{Jihyun
  Kim}, {and} \bibinfo{person}{Howon Kim}.} \bibinfo{year}{2017}\natexlab{}.
\newblock \showarticletitle{An Effective Intrusion Detection Classifier Using
  Long Short-Term Memory with Gradient Descent Optimization}. In
  \bibinfo{booktitle}{\emph{IEEE 2017 International Conference on Platform
  Technology and Service (PlatCon) - Busan, South Korea}}.
  \bibinfo{pages}{1--6}.
\newblock


\bibitem[\protect\citeauthoryear{Lee, Sugiyama, von Luxburg, Guyon, and
  Garnett}{Lee et~al\mbox{.}}{2016}]%
        {DBLPnips2016}
\bibfield{editor}{\bibinfo{person}{Daniel~D. Lee}, \bibinfo{person}{Masashi
  Sugiyama}, \bibinfo{person}{Ulrike von Luxburg}, \bibinfo{person}{Isabelle
  Guyon}, {and} \bibinfo{person}{Roman Garnett}} (Eds.).
  \bibinfo{year}{2016}\natexlab{}.
\newblock \bibinfo{booktitle}{\emph{Advances in Neural Information Processing
  Systems 29: Annual Conference on Neural Information Processing Systems 2016,
  December 5-10, 2016, Barcelona, Spain}}.
\newblock
\urldef\tempurl%
\url{http://papers.nips.cc/book/advances-in-neural-information-processing-systems-29-2016}
\showURL{%
\tempurl}


\bibitem[\protect\citeauthoryear{Liao, Lin, Lin, and Tung}{Liao
  et~al\mbox{.}}{2013}]%
        {DBLP123}
\bibfield{author}{\bibinfo{person}{Hung{-}Jen Liao},
  \bibinfo{person}{Chun{-}Hung~Richard Lin}, \bibinfo{person}{Ying{-}Chih Lin},
  {and} \bibinfo{person}{Kuang{-}Yuan Tung}.} \bibinfo{year}{2013}\natexlab{}.
\newblock \showarticletitle{Intrusion detection system: {A} comprehensive
  review}.
\newblock \bibinfo{journal}{\emph{J. Netw. Comput. Appl.}}
  \bibinfo{volume}{36}, \bibinfo{number}{1} (\bibinfo{year}{2013}),
  \bibinfo{pages}{16--24}.
\newblock
\urldef\tempurl%
\url{https://doi.org/10.1016/j.jnca.2012.09.004}
\showDOI{\tempurl}


\bibitem[\protect\citeauthoryear{Lillicrap, Hunt, Pritzel, Heess, Erez, Tassa,
  Silver, and Wierstra}{Lillicrap et~al\mbox{.}}{2015}]%
        {lillicrap2015}
\bibfield{author}{\bibinfo{person}{Timothy~P Lillicrap},
  \bibinfo{person}{Jonathan~J Hunt}, \bibinfo{person}{Alexander Pritzel},
  \bibinfo{person}{Nicolas Heess}, \bibinfo{person}{Tom Erez},
  \bibinfo{person}{Yuval Tassa}, \bibinfo{person}{David Silver}, {and}
  \bibinfo{person}{Daan Wierstra}.} \bibinfo{year}{2015}\natexlab{}.
\newblock \showarticletitle{Continuous control with deep reinforcement
  learning}.
\newblock \bibinfo{journal}{\emph{arXiv preprint arXiv:1509.02971}}
  (\bibinfo{year}{2015}).
\newblock


\bibitem[\protect\citeauthoryear{Naseer, Saleem, Khalid, Bashir, Han, Iqbal,
  and Han}{Naseer et~al\mbox{.}}{2018}]%
        {DBLPSKBHIH18}
\bibfield{author}{\bibinfo{person}{Sheraz Naseer}, \bibinfo{person}{Yasir
  Saleem}, \bibinfo{person}{Shehzad Khalid}, \bibinfo{person}{Muhammad~Khawar
  Bashir}, \bibinfo{person}{Jihun Han}, \bibinfo{person}{Muhammad~Munwar
  Iqbal}, {and} \bibinfo{person}{Kijun Han}.} \bibinfo{year}{2018}\natexlab{}.
\newblock \showarticletitle{Enhanced Network Anomaly Detection Based on Deep
  Neural Networks}.
\newblock \bibinfo{journal}{\emph{{IEEE} Access}}  \bibinfo{volume}{6}
  (\bibinfo{year}{2018}), \bibinfo{pages}{48231--48246}.
\newblock
\urldef\tempurl%
\url{https://doi.org/10.1109/ACCESS.2018.2863036}
\showDOI{\tempurl}


\bibitem[\protect\citeauthoryear{Qu, Zhang, Shao, and Qi}{Qu
  et~al\mbox{.}}{2017}]%
        {DBLPQuZSQ17}
\bibfield{author}{\bibinfo{person}{Feng Qu}, \bibinfo{person}{Jitao Zhang},
  \bibinfo{person}{Zetian Shao}, {and} \bibinfo{person}{Shuzhuang Qi}.}
  \bibinfo{year}{2017}\natexlab{}.
\newblock \showarticletitle{An Intrusion Detection Model Based on Deep Belief
  Network}. In \bibinfo{booktitle}{\emph{Proceedings of the {VI} International
  Conference on Network, Communication and Computingm, {ICNCC} 2017, Kunming,
  China, December 8-10, 2017}}. \bibinfo{pages}{97--101}.
\newblock
\urldef\tempurl%
\url{https://doi.org/10.1145/3171592.3171598}
\showDOI{\tempurl}


\bibitem[\protect\citeauthoryear{Rajkumar, Lee, Sha, and Stankovic}{Rajkumar
  et~al\mbox{.}}{2010}]%
        {DBLPSS10}
\bibfield{author}{\bibinfo{person}{Ragunathan Rajkumar}, \bibinfo{person}{Insup
  Lee}, \bibinfo{person}{Lui Sha}, {and} \bibinfo{person}{John~A. Stankovic}.}
  \bibinfo{year}{2010}\natexlab{}.
\newblock \showarticletitle{Cyber-physical systems: the next computing
  revolution}. In \bibinfo{booktitle}{\emph{Proceedings of the 47th Design
  Automation Conference, {DAC} 2010, Anaheim, California, USA, July 13-18,
  2010}}. \bibinfo{pages}{731--736}.
\newblock
\urldef\tempurl%
\url{https://doi.org/10.1145/1837274.1837461}
\showDOI{\tempurl}


\bibitem[\protect\citeauthoryear{Shone, Tran, Phai, and Shi}{Shone
  et~al\mbox{.}}{2018}]%
        {DBLPTPS18}
\bibfield{author}{\bibinfo{person}{Nathan Shone}, \bibinfo{person}{Nguyen~Ngoc
  Tran}, \bibinfo{person}{Vu~Dinh Phai}, {and} \bibinfo{person}{Qi Shi}.}
  \bibinfo{year}{2018}\natexlab{}.
\newblock \showarticletitle{A Deep Learning Approach to Network Intrusion
  Detection}.
\newblock \bibinfo{journal}{\emph{{IEEE} Trans. Emerging Topics in Comput.
  Intellig.}} \bibinfo{volume}{2}, \bibinfo{number}{1} (\bibinfo{year}{2018}),
  \bibinfo{pages}{41--50}.
\newblock
\urldef\tempurl%
\url{https://doi.org/10.1109/TETCI.2017.2772792}
\showDOI{\tempurl}


\bibitem[\protect\citeauthoryear{Silver, Lever, Heess, Degris, Wierstra, and
  Riedmiller}{Silver et~al\mbox{.}}{2014}]%
        {Silverarticle}
\bibfield{author}{\bibinfo{person}{David Silver}, \bibinfo{person}{Guy Lever},
  \bibinfo{person}{Nicolas Heess}, \bibinfo{person}{Thomas Degris},
  \bibinfo{person}{Daan Wierstra}, {and} \bibinfo{person}{Martin Riedmiller}.}
  \bibinfo{year}{2014}\natexlab{}.
\newblock \showarticletitle{Deterministic Policy Gradient Algorithms}.
\newblock \bibinfo{journal}{\emph{31st International Conference on Machine
  Learning, ICML 2014}}  \bibinfo{volume}{1} (\bibinfo{date}{06}
  \bibinfo{year}{2014}).
\newblock


\bibitem[\protect\citeauthoryear{Sutton and Barto}{Sutton and Barto}{1998}]%
        {DBLPSutton98}
\bibfield{author}{\bibinfo{person}{Richard~S. Sutton} {and}
  \bibinfo{person}{Andrew~G. Barto}.} \bibinfo{year}{1998}\natexlab{}.
\newblock \showarticletitle{Reinforcement Learning: An Introduction}.
\newblock \bibinfo{journal}{\emph{{IEEE} Trans. Neural Networks}}
  \bibinfo{volume}{9}, \bibinfo{number}{5} (\bibinfo{year}{1998}),
  \bibinfo{pages}{1054--1054}.
\newblock
\urldef\tempurl%
\url{https://doi.org/10.1109/TNN.1998.712192}
\showDOI{\tempurl}


\bibitem[\protect\citeauthoryear{Xu, Chen, Zhang, and Wu}{Xu
  et~al\mbox{.}}{2017}]%
        {Xu2017Incremental}
\bibfield{author}{\bibinfo{person}{Binhan Xu}, \bibinfo{person}{Shuyu Chen},
  \bibinfo{person}{Hancui Zhang}, {and} \bibinfo{person}{Tianshu Wu}.}
  \bibinfo{year}{2017}\natexlab{}.
\newblock \showarticletitle{Incremental k-NN SVM method in intrusion
  detection}. In \bibinfo{booktitle}{\emph{2017 8th IEEE International
  Conference on Software Engineering and Service Science (ICSESS)}}.
\newblock


\bibitem[\protect\citeauthoryear{Yao, Jiang, and Qian}{Yao
  et~al\mbox{.}}{2019}]%
        {inbook009}
\bibfield{author}{\bibinfo{person}{Haipeng Yao}, \bibinfo{person}{Chunxiao
  Jiang}, {and} \bibinfo{person}{Yi Qian}.} \bibinfo{year}{2019}\natexlab{}.
\newblock \bibinfo{booktitle}{\emph{Intelligent Network Awareness}}.
\newblock
\showISBNx{978-3-030-15028-0}
\urldef\tempurl%
\url{https://doi.org/10.1007/978-3-030-15028-0_3}
\showDOI{\tempurl}


\bibitem[\protect\citeauthoryear{Yin, Zhu, Fei, and He}{Yin
  et~al\mbox{.}}{2017}]%
        {DBLPZFH17}
\bibfield{author}{\bibinfo{person}{Chuanlong Yin}, \bibinfo{person}{Yuefei
  Zhu}, \bibinfo{person}{Jinlong Fei}, {and} \bibinfo{person}{Xin{-}Zheng He}.}
  \bibinfo{year}{2017}\natexlab{}.
\newblock \showarticletitle{A Deep Learning Approach for Intrusion Detection
  Using Recurrent Neural Networks}.
\newblock \bibinfo{journal}{\emph{{IEEE} Access}}  \bibinfo{volume}{5}
  (\bibinfo{year}{2017}), \bibinfo{pages}{21954--21961}.
\newblock
\urldef\tempurl%
\url{https://doi.org/10.1109/ACCESS.2017.2762418}
\showDOI{\tempurl}


\bibitem[\protect\citeauthoryear{Zhang, Zulkernine, and Haque}{Zhang
  et~al\mbox{.}}{2008}]%
        {DBLPZH08}
\bibfield{author}{\bibinfo{person}{Jiong Zhang}, \bibinfo{person}{Mohammad
  Zulkernine}, {and} \bibinfo{person}{Anwar Haque}.}
  \bibinfo{year}{2008}\natexlab{}.
\newblock \showarticletitle{Random-Forests-Based Network Intrusion Detection
  Systems}.
\newblock \bibinfo{journal}{\emph{{IEEE} Trans. Systems, Man, and Cybernetics,
  Part {C}}} \bibinfo{volume}{38}, \bibinfo{number}{5} (\bibinfo{year}{2008}),
  \bibinfo{pages}{649--659}.
\newblock
\urldef\tempurl%
\url{https://doi.org/10.1109/TSMCC.2008.923876}
\showDOI{\tempurl}


\end{thebibliography}


\begin{thebibliography}{44}
\providecommand{\natexlab}[1]{#1}
\providecommand{\url}[1]{\texttt{#1}}
\expandafter\ifx\csname urlstyle\endcsname\relax
  \providecommand{\doi}[1]{doi: #1}\else
  \providecommand{\doi}{doi: \begingroup \urlstyle{rm}\Url}\fi

\bibitem[Addanki et~al.(2019)Addanki, Venkatakrishnan, Gupta, Mao, and
  Alizadeh]{Placeto18}
Addanki, R., Venkatakrishnan, S.~B., Gupta, S., Mao, H., and Alizadeh, M.
\newblock Placeto: Learning generalizable device placement algorithms for
  distributed machine learning.
\newblock \emph{CoRR}, abs/1906.08879, 2019.
\newblock URL \url{http://arxiv.org/abs/1906.08879}.

\bibitem[Agnihotri et~al.(2005)Agnihotri, Ono, and Madden]{fengshui2005}
Agnihotri, A., Ono, S., and Madden, P.
\newblock Recursive bisection placement: Feng shui 5.0 implementation details.
\newblock In \emph{Proceedings of the International Symposium on Physical
  Design}, pp.\  230--232, 01 2005.
\newblock \doi{10.1145/1055137.1055186}.

\bibitem[{Bo Hu} \& {Marek-Sadowska}(2005){Bo Hu} and
  {Marek-Sadowska}]{mfar2005}
{Bo Hu} and {Marek-Sadowska}, M.
\newblock Multilevel fixed-point-addition-based vlsi placement.
\newblock \emph{IEEE Transactions on Computer-Aided Design of Integrated
  Circuits and Systems}, 24\penalty0 (8):\penalty0 1188--1203, Aug 2005.
\newblock ISSN 1937-4151.
\newblock \doi{10.1109/TCAD.2005.850802}.

\bibitem[Brenner et~al.(2008)Brenner, Struzyna, and Vygen]{bonnplace2008}
Brenner, U., Struzyna, M., and Vygen, J.
\newblock Bonnplace: Placement of leading-edge chips by advanced combinatorial
  algorithms.
\newblock \emph{Trans. Comp.-Aided Des. Integ. Cir. Sys.}, 27\penalty0
  (9):\penalty0 1607–1620, September 2008.
\newblock ISSN 0278-0070.
\newblock \doi{10.1109/TCAD.2008.927674}.
\newblock URL \url{https://doi.org/10.1109/TCAD.2008.927674}.

\bibitem[Breuer(1977)]{MinCutBreuer1977}
Breuer, M.~A.
\newblock A class of min-cut placement algorithms.
\newblock In \emph{Proceedings of the 14th Design Automation Conference}, DAC
  ’77, pp.\  284–290. IEEE Press, 1977.

\bibitem[{Chen} et~al.(2008){Chen}, {Jiang}, {Hsu}, {Chen}, and
  {Chang}]{ntuplace32008}
{Chen}, T., {Jiang}, Z., {Hsu}, T., {Chen}, H., and {Chang}, Y.
\newblock Ntuplace3: An analytical placer for large-scale mixed-size designs
  with preplaced blocks and density constraints.
\newblock \emph{IEEE Transactions on Computer-Aided Design of Integrated
  Circuits and Systems}, 27\penalty0 (7):\penalty0 1228--1240, July 2008.
\newblock ISSN 1937-4151.
\newblock \doi{10.1109/TCAD.2008.923063}.

\bibitem[Chen et~al.(2006)Chen, Jiang, Hsu, Chen, and Chang]{NTUPlacer06}
Chen, T.-C., Jiang, Z.-W., Hsu, T.-C., Chen, H.-C., and Chang, Y.-W.
\newblock A high-quality mixed-size analytical placer considering preplaced
  blocks and density constraints.
\newblock In \emph{Proceedings of the 2006 IEEE/ACM International Conference on
  Computer-Aided Design}, ICCAD ’06, pp.\  187–192, New York, NY, USA,
  2006. Association for Computing Machinery.
\newblock ISBN 1595933891.

\bibitem[{Cheng} et~al.(2019){Cheng}, {Kahng}, {Kang}, and {Wang}]{RePlAce19}
{Cheng}, C., {Kahng}, A.~B., {Kang}, I., and {Wang}, L.
\newblock Replace: Advancing solution quality and routability validation in
  global placement.
\newblock \emph{IEEE Transactions on Computer-Aided Design of Integrated
  Circuits and Systems}, 38\penalty0 (9):\penalty0 1717--1730, 2019.

\bibitem[{Chung-Kuan Cheng} \& {Kuh}(1984){Chung-Kuan Cheng} and
  {Kuh}]{ResistiveNetwork1984}
{Chung-Kuan Cheng} and {Kuh}, E.~S.
\newblock Module placement based on resistive network optimization.
\newblock \emph{IEEE Transactions on Computer-Aided Design of Integrated
  Circuits and Systems}, 3\penalty0 (3):\penalty0 218--225, July 1984.
\newblock ISSN 1937-4151.
\newblock \doi{10.1109/TCAD.1984.1270078}.

\bibitem[{Fiduccia} \& {Mattheyses}(1982){Fiduccia} and
  {Mattheyses}]{fiduccia1982}
{Fiduccia}, C.~M. and {Mattheyses}, R.~M.
\newblock A linear-time heuristic for improving network partitions.
\newblock In \emph{19th Design Automation Conference}, pp.\  175--181, June
  1982.
\newblock \doi{10.1109/DAC.1982.1585498}.

\bibitem[Gilbert \& Pollak(1968)Gilbert and Pollak]{gilbert1968steiner}
Gilbert, E.~N. and Pollak, H.~O.
\newblock Steiner minimal trees.
\newblock \emph{SIAM Journal on Applied Mathematics}, 16\penalty0 (1):\penalty0
  1--29, 1968.

\bibitem[Hanan \& Kurtzberg(1972)Hanan and Kurtzberg]{forcedirected1972}
Hanan, M. and Kurtzberg, J.
\newblock Placement techniques.
\newblock In \emph{Design Automation of Digital Systems}, 1972.

\bibitem[{Hsu} et~al.(2011){Hsu}, {Chang}, and
  {Balabanov}]{weightedaverage2011}
{Hsu}, M., {Chang}, Y., and {Balabanov}, V.
\newblock Tsv-aware analytical placement for 3d ic designs.
\newblock In \emph{2011 48th ACM/EDAC/IEEE Design Automation Conference (DAC)},
  pp.\  664--669, June 2011.

\bibitem[Huang et~al.(2019)Huang, Xie, Fang, Yu, Ren, Fang, Chen, and
  Hu]{cnnplacer19}
Huang, Y., Xie, Z., Fang, G., Yu, T., Ren, H., Fang, S., Chen, Y., and Hu, J.
\newblock Routability-driven macro placement with embedded cnn-based prediction
  model.
\newblock In Teich, J. and Fummi, F. (eds.), \emph{Design, Automation {\&} Test
  in Europe Conference {\&} Exhibition, {DATE} 2019, Florence, Italy, March
  25-29, 2019}, pp.\  180--185. {IEEE}, 2019.

\bibitem[{Kahng} et~al.(2005){Kahng}, {Reda}, and {Qinke Wang}]{aplace22005}
{Kahng}, A.~B., {Reda}, S., and {Qinke Wang}.
\newblock Architecture and details of a high quality, large-scale analytical
  placer.
\newblock In \emph{ICCAD-2005. IEEE/ACM International Conference on
  Computer-Aided Design, 2005.}, pp.\  891--898, Nov 2005.
\newblock \doi{10.1109/ICCAD.2005.1560188}.

\bibitem[Karypis \& Kumar(1998)Karypis and Kumar]{hmetis1998}
Karypis, G. and Kumar, V.
\newblock A hypergraph partitioning package.
\newblock In \emph{HMETIS}, 1998.

\bibitem[Kernighan(1985)]{TerminalPropagation1985}
Kernighan, D.~.
\newblock A procedure for placement of standard-cell vlsi circuits.
\newblock In \emph{IEEE TCAD}, 1985.

\bibitem[{Kim} \& {Markov}(2012){Kim} and {Markov}]{complx2012}
{Kim}, M. and {Markov}, I.~L.
\newblock Complx: A competitive primal-dual lagrange optimization for global
  placement.
\newblock In \emph{DAC Design Automation Conference 2012}, pp.\  747--755, June
  2012.

\bibitem[Kim et~al.(2010)Kim, Lee, and Markov]{simpl2010}
Kim, M.-C., Lee, D.-J., and Markov, I.~L.
\newblock Simpl: An effective placement algorithm.
\newblock In \emph{Proceedings of the International Conference on
  Computer-Aided Design}, ICCAD ’10, pp.\  649–656. IEEE Press, 2010.
\newblock ISBN 9781424481927.

\bibitem[Kim et~al.(2012{\natexlab{a}})Kim, Viswanathan, Alpert, Markov, and
  Ramji]{MAPLE12}
Kim, M.-C., Viswanathan, N., Alpert, C.~J., Markov, I.~L., and Ramji, S.
\newblock Maple: Multilevel adaptive placement for mixed-size designs.
\newblock In \emph{Proceedings of the 2012 ACM International Symposium on
  International Symposium on Physical Design}, ISPD, pp.\  193–200, New York,
  NY, USA, 2012{\natexlab{a}}. Association for Computing Machinery.

\bibitem[Kim et~al.(2012{\natexlab{b}})Kim, Viswanathan, Alpert, Markov, and
  Ramji]{maple2012}
Kim, M.-C., Viswanathan, N., Alpert, C.~J., Markov, I.~L., and Ramji, S.
\newblock Maple: Multilevel adaptive placement for mixed-size designs.
\newblock In \emph{Proceedings of the 2012 ACM International Symposium on
  International Symposium on Physical Design}, ISPD ’12, pp.\  193–200, New
  York, NY, USA, 2012{\natexlab{b}}. Association for Computing Machinery.
\newblock ISBN 9781450311670.
\newblock \doi{10.1145/2160916.2160958}.
\newblock URL \url{https://doi.org/10.1145/2160916.2160958}.

\bibitem[Kirkpatrick et~al.(1983)Kirkpatrick, Gelatt, and
  Vecchi]{SimulatedAnnealing}
Kirkpatrick, S., Gelatt, C.~D., and Vecchi, M.~P.
\newblock Optimization by simulated annealing.
\newblock \emph{Science}, 220\penalty0 (4598):\penalty0 671--680, 1983.
\newblock ISSN 0036-8075.
\newblock \doi{10.1126/science.220.4598.671}.
\newblock URL \url{https://science.sciencemag.org/content/220/4598/671}.

\bibitem[Lin et~al.(2013)Lin, Chu, Shinnerl, Bustany, and Nedelchev]{polar2013}
Lin, T., Chu, C., Shinnerl, J.~R., Bustany, I., and Nedelchev, I.
\newblock Polar: Placement based on novel rough legalization and refinement.
\newblock In \emph{Proceedings of the International Conference on
  Computer-Aided Design}, ICCAD ’13, pp.\  357–362. IEEE Press, 2013.
\newblock ISBN 9781479910694.

\bibitem[Lin et~al.(2019)Lin, Dhar, Li, Ren, Khailany, and Pan]{Dreamplace19}
Lin, Y., Dhar, S., Li, W., Ren, H., Khailany, B., and Pan, D.~Z.
\newblock Dreamplace: Deep learning toolkit-enabled gpu acceleration for modern
  vlsi placement.
\newblock In \emph{Proceedings of the 56th Annual Design Automation Conference
  2019}, DAC ’19, 2019.

\bibitem[Lu et~al.(2015)Lu, Chen, Chang, Sha, Huang, Teng, and Cheng]{EPlace15}
Lu, J., Chen, P., Chang, C.-C., Sha, L., Huang, D. J.-H., Teng, C.-C., and
  Cheng, C.-K.
\newblock Eplace: Electrostatics-based placement using fast fourier transform
  and nesterov’s method.
\newblock \emph{ACM Trans. Des. Autom. Electron. Syst.}, 20\penalty0 (2), 2015.
\newblock ISSN 1084-4309.

\bibitem[{Lu} et~al.(2015){Lu}, {Zhuang}, {Chen}, {Chang}, {Chang}, {Wong},
  {Sha}, {Huang}, {Luo}, {Teng}, and {Cheng}]{EPlacemixed15}
{Lu}, J., {Zhuang}, H., {Chen}, P., {Chang}, H., {Chang}, C., {Wong}, Y.,
  {Sha}, L., {Huang}, D., {Luo}, Y., {Teng}, C., and {Cheng}, C.
\newblock eplace-ms: Electrostatics-based placement for mixed-size circuits.
\newblock \emph{IEEE Transactions on Computer-Aided Design of Integrated
  Circuits and Systems}, 34\penalty0 (5):\penalty0 685--698, 2015.

\bibitem[Lu et~al.(2016)Lu, Zhuang, Kang, Chen, and Cheng]{EPLace3D16}
Lu, J., Zhuang, H., Kang, I., Chen, P., and Cheng, C.-K.
\newblock Eplace-3d: Electrostatics based placement for 3d-ics.
\newblock In \emph{Proceedings of the 2016 on International Symposium on
  Physical Design}, ISPD ’16, New York, NY, USA, 2016. Association for
  Computing Machinery.
\newblock ISBN 9781450340397.
\newblock \doi{10.1145/2872334.2872361}.
\newblock URL \url{https://doi.org/10.1145/2872334.2872361}.

\bibitem[Nazi et~al.(2019)Nazi, Hang, Goldie, Ravi, and
  Mirhoseini]{nazi2019gap}
Nazi, A., Hang, W., Goldie, A., Ravi, S., and Mirhoseini, A.
\newblock Gap: Generalizable approximate graph partitioning framework, 2019.

\bibitem[Obermeier et~al.(2005)Obermeier, Ranke, and Johannes]{kraftwerk2005}
Obermeier, B., Ranke, H., and Johannes, F.
\newblock Kraftwerk: a versatile placement approach.
\newblock In \emph{ISPD}, pp.\  242--244, 01 2005.
\newblock \doi{10.1145/1055137.1055190}.

\bibitem[Paliwal et~al.(2019)Paliwal, Gimeno, Nair, Li, Lubin, Kohli, and
  Vinyals]{REGAL19}
Paliwal, A.~S., Gimeno, F., Nair, V., Li, Y., Lubin, M., Kohli, P., and
  Vinyals, O.
\newblock Regal: Transfer learning for fast optimization of computation graphs.
\newblock \emph{ArXiv}, abs/1905.02494, 2019.

\bibitem[{Ren-Song Tsay} et~al.(1988){Ren-Song Tsay}, {Kuh}, and {Chi-Ping
  Hsu}]{proud1988}
{Ren-Song Tsay}, {Kuh}, E.~S., and {Chi-Ping Hsu}.
\newblock Proud: a sea-of-gates placement algorithm.
\newblock \emph{IEEE Design Test of Computers}, 5\penalty0 (6):\penalty0
  44--56, Dec 1988.
\newblock ISSN 1558-1918.
\newblock \doi{10.1109/54.9271}.

\bibitem[Roy et~al.(2007)Roy, Papa, and Markov]{capo2007}
Roy, J.~A., Papa, D.~A., and Markov, I.~L.
\newblock \emph{Capo: Congestion-Driven Placement for Standard-cell and RTL
  Netlists with Incremental Capability}, pp.\  97--133.
\newblock Springer US, Boston, MA, 2007.

\bibitem[Sarrafzadeh et~al.(2003)Sarrafzadeh, Wang, and Yang]{dragon}
Sarrafzadeh, M., Wang, M., and Yang, X.
\newblock \emph{Dragon: A Placement Framework}, pp.\  57--89.
\newblock Springer, 01 2003.
\newblock ISBN 978-1-4419-5309-4.
\newblock \doi{10.1007/978-1-4757-3781-3_3}.

\bibitem[Schulman et~al.(2017)Schulman, Wolski, Dhariwal, Radford, and
  Klimov]{ppo17}
Schulman, J., Wolski, F., Dhariwal, P., Radford, A., and Klimov, O.
\newblock Proximal policy optimization algorithms, 2017.

\bibitem[Sechen \& Sangiovanni-Vincentelli(1986)Sechen and
  Sangiovanni-Vincentelli]{DAC-1986-SechenS}
Sechen, C. and Sangiovanni-Vincentelli, A.~L.
\newblock {TimberWolf3.2: a new standard cell placement and global routing
  package}.
\newblock In \emph{{DAC}}, pp.\  432--439. {IEEE Computer Society Press}, 1986.
\newblock \doi{10.1145/318013.318083}.

\bibitem[Shahookar \& Mazumder(1991)Shahookar and Mazumder]{hpwl1991}
Shahookar, K. and Mazumder, P.
\newblock Vlsi cell placement techniques.
\newblock \emph{ACM Comput. Surv.}, 23\penalty0 (2):\penalty0 143–220, June
  1991.
\newblock ISSN 0360-0300.
\newblock \doi{10.1145/103724.103725}.
\newblock URL \url{https://doi.org/10.1145/103724.103725}.

\bibitem[{Spindler} et~al.(2008){Spindler}, {Schlichtmann}, and
  {Johannes}]{kraftwerk22008}
{Spindler}, P., {Schlichtmann}, U., and {Johannes}, F.~M.
\newblock Kraftwerk2—a fast force-directed quadratic placement approach using
  an accurate net model.
\newblock \emph{IEEE Transactions on Computer-Aided Design of Integrated
  Circuits and Systems}, 27\penalty0 (8):\penalty0 1398--1411, Aug 2008.
\newblock ISSN 1937-4151.
\newblock \doi{10.1109/TCAD.2008.925783}.

\bibitem[{Tao Luo} \& {Pan}(2008){Tao Luo} and {Pan}]{dplace2008}
{Tao Luo} and {Pan}, D.~Z.
\newblock Dplace2.0: A stable and efficient analytical placement based on
  diffusion.
\newblock In \emph{2008 Asia and South Pacific Design Automation Conference},
  pp.\  346--351, March 2008.
\newblock \doi{10.1109/ASPDAC.2008.4483972}.

\bibitem[Viswanathan et~al.(2007{\natexlab{a}})Viswanathan, Nam, Alpert,
  Villarrubia, Ren, and Chu]{rql2007}
Viswanathan, N., Nam, G.-J., Alpert, C., Villarrubia, P., Ren, H., and Chu, C.
\newblock Rql: Global placement via relaxed quadratic spreading and
  linearization.
\newblock In \emph{Proceedings - Design Automation Conference}, pp.\  453--458,
  07 2007{\natexlab{a}}.
\newblock ISBN 978-1-59593-627-1.
\newblock \doi{10.1145/1278480.1278599}.

\bibitem[Viswanathan et~al.(2007{\natexlab{b}})Viswanathan, Pan, and
  Chu]{fastplace2007}
Viswanathan, N., Pan, M., and Chu, C.
\newblock \emph{FastPlace: An Efficient Multilevel Force-Directed Placement
  Algorithm}, pp.\  193--228.
\newblock Springer, 01 2007{\natexlab{b}}.
\newblock \doi{10.1007/978-0-387-68739-1_8}.

\bibitem[William et~al.(2001)William, Ross, and Lu]{logsumexp2001}
William, N., Ross, D., and Lu, S.
\newblock Non-linear optimization system and method for wire length and delay
  optimization for an automatic electric circuit placer.
\newblock In \emph{Patent}, 2001.

\bibitem[Zhang \& Chen(2018)Zhang and Chen]{zhang2018link}
Zhang, M. and Chen, Y.
\newblock Link prediction based on graph neural networks, 2018.

\bibitem[Zhiyao Xie Duke~Univeristy(2018)]{RouteNet18}
Zhiyao Xie Duke~Univeristy, Durham, N. U. . Y.-H. H. . G.-Q. F. . H. R. . S.-Y.
  F. . Y. C. . J.~H.
\newblock Routenet: Routability prediction for mixed-size designs using
  convolutional neural network.
\newblock In \emph{IEEE/ACM International Conference on Computer-Aided Design
  (ICCAD}, 2018.

\bibitem[Zhou et~al.(2019)Zhou, Roy, Abdolrashidi, Wong, Ma, Xu, Zhong, Liu,
  Goldie, Mirhoseini, and Laudon]{zhou2019gdp}
Zhou, Y., Roy, S., Abdolrashidi, A., Wong, D., Ma, P.~C., Xu, Q., Zhong, M.,
  Liu, H., Goldie, A., Mirhoseini, A., and Laudon, J.
\newblock Gdp: Generalized device placement for dataflow graphs, 2019.

\end{thebibliography}
\bibliographystyle{ACM-Reference-Format}
\end{document}